# Tensor Interaction in SkHF theory and its influence on evolution of nuclear shells

Rupayan Bhattacharya*
University of Calcutta

Using Skyrme's density dependent interaction with SKP force parameters the evolution of nuclear shells has been studied in Hartree-Fock calculations. The effect of tensor interaction optimization has been done in reproducing the observed splitting of shell model states of $^{40,48}$Ca, $^{56}$Ni and $^{208}$Pb. Spin-orbit splitting in Ca-isotopes, $^{56}$Ni, $^{90}$Zr, N = 82 isotones, Sn-isotopes and evolution of gaps in Z, N = 8, 20 have been reanalyzed with the inclusion of tensor interaction. For doubly shell closed nuclei it has been observed that tensor interaction is sensitive to spin saturation of nuclear shells.

---

*rup_bhat@hotmail.com

# I. INTRODUCTION

The important role of the tensor component of the nucleon-nucleon interaction in the splitting and regrouping of the spin-orbit doublets of single particle spectrum leading to formation of shells beyond the conventional ones was pointed out in the seventies [1 – 6] but it was not included in meanfield structure calculations. It has drawn lots of attention after appearance of several interesting works in recent times. Advent of new accelerators enabled scientists to investigate the properties and structures of nuclei far from the stability line. New experiments with radioactive particle beams revealed fascinating features like halos and skins [7 - 9], new areas of magicity [10] and pygmy dipole resonance [11,12]. Effect of tensor interaction on the structure of doubly shell closed nucleus $^{16}O$ and the utility of the tensor force in the evolution of shell structure in proton (-neutron) rich nuclei through shell model have been shown by many authors [13 – 17]. Study of shell evolution and modification of magic numbers in exotic nuclei after inclusion of the tensor component of the nucleon-nucleon interaction in the self-consistent mean field calculations generated interest in the nuclear structure study circle [18 – 24].

Zheng and Zamick [13] had constructed a generalized interaction of the form $V_c + XV_{so} + yV_t$. By varying x and y they could determine the effects of the two-body spin-orbit and tensor interactions in nuclei. When applied to a closed LS shell like $^{16}O$, they found that the isoscalar states are more affected than the isovector states. In the work of Otsuka et al [14-17] a Gaussian tensor force was added to a standard form of Gogny force after adjustment of the long range part of a one $\pi-$ and $\rho-$ exchange potential. The relative movement of some single particle levels was explained at the cost of absolute separation in energy. For reproduction of single particle spectrum near the Fermi surface Brown et al [25] have optimized a Skyrme interaction in the Hartree-Fock formalism with an additional ingredient of zero range tensor force. However, their attempt could not describe correctly the l-dependence of the spin orbit splitting. In order to emphasise the importance of tensor force in evolution of shells Colo et al [18] and Brink and Stancu [19] introduced Skyrme's tensor force perturbatively in their calculation of energy difference of shell model states in Z=50 isotopes and N = 82 isotones using Sly5 [26] and SIII [2] parameters respectively. Lesinsky et al [20] have shown in great details how the presence of a tensor force leads to a strong rearrangement of spin-orbit force and they have generated a large number of force parameters conforming to the needs of the situation. They found that the variation of the spin-orbit force has a larger impact on the systematic of single particle spectra than the tensor terms themselves. Furthermore, the main effect of the tensor force lies in the evolution of the spin-orbit splitting with N and Z. Using a strategy of shifting the focus from ground-state bulk properties (e.g. total nuclear mass, r.m.s. charge radii) to single particle properties of shell closed nuclei Zalewski et al [21] have shown exhaustively that a change is required in the isoscalar spin-orbit strength and the tensor coupling constants for reproduction of the single particle spectrum. Tensor effect in shell evolution has also been addressed by Moreno-Torres et al [22] where they have suggested that the magic numbers 8 and 20 are suitable for fitting the tensor parameters in mean-field approach. Zou et al have shown that inclusion of tensor terms can fairly well explain the trend of single particle states of Ca isotopes and can qualitatively explain the reduction of the neutron spin-orbit splitting as one moves from $^{48}Ca$ to $^{46}Ar$. Wang et al [24] have investigated the evolution of gap between the states just below and just above the Fermi surface of nuclei having Z, N = 8, 20, 28 after inclusion of tensor forces. The experimental trend in gap evolution is better reproduced in tensor component included HFB calculations.

In a seminal work Bender et al [27] have shown that the addition of only a tensor term in the spin-orbit part perturbatively, like for SLy5+T, to an existing parametrization will modify all contributions to the mean fields and the energy. The results may be unstable as perturbative modification of a well-adjusted parametrization might spoil its predictive power in unexpected ways. The tensor and spin-orbit contributions to the total energy and to the spin-orbit fields are tightly interwoven. It is well known that constraining the tensor contribution in a relatively small region of the nuclear chart might be misleading when aiming at a universal functional.

In order to study the evolution of shells after incorporating the effect of tensor force in Skyrme–Hartree–Fock theory we have studied the spin-orbit splitting of some shell model states of doubly shell closed nuclei, shell gap evolutions of Z, N = 8, 20 nuclei, difference in single particle(-hole) energy levels near the Fermi surfaces of Z = 50 isotopes, Ca – isotopes and N = 82 isotones. After comparing the performances of several set of force parameters we have used SKP [28] set of parameters, which represents a zero-range interaction for our calculations. For pairing interaction we have used standard BCS prescription. The paper is organised as follows: in the Sec. II the theoretical frame work for our calculation has been introduced. The results and corresponding discussions are given in Sec. III. Summary and conclusion complete the structure of the paper.

## II. THEORETICAL FORMULATION

The tensor interaction in HF theory is given by

$$V_T = \frac{T}{2}\left\{\left[(\boldsymbol{\sigma_1}\cdot\boldsymbol{k'})(\boldsymbol{\sigma_2}\cdot\boldsymbol{k'}) - \frac{1}{3}(\boldsymbol{\sigma_1}\cdot\boldsymbol{\sigma_2})k'^2\right]\delta(\boldsymbol{r_1}-\boldsymbol{r_2}) + \delta(\boldsymbol{r_1}-\boldsymbol{r_2})\left[(\boldsymbol{\sigma_1}\cdot\boldsymbol{k})(\boldsymbol{\sigma_2}\cdot\boldsymbol{k}) - \frac{1}{3}(\boldsymbol{\sigma_1}\cdot\boldsymbol{\sigma_2})k^2\right]\right\}$$

$$+ U\{(\boldsymbol{\sigma_1}\cdot\boldsymbol{k'})\delta(\boldsymbol{r_1}-\boldsymbol{r_2})(\boldsymbol{\sigma_1}\cdot\boldsymbol{k}) - \frac{1}{3}(\boldsymbol{\sigma_1}\cdot\boldsymbol{\sigma_2})\times[\boldsymbol{k'}\cdot\delta(r_1-r_2)\boldsymbol{k}]\} \qquad (1)$$

As we are interested in the tensor interaction and its effect, we fix our attention on the time-even tensor and spin-orbit parts of the energy density function,

$$H_T = C_0^J J_0^2 + C_1^J J_1^2 \qquad (2)$$

$$H_{SO} = C_0^{\nabla J} \rho_0 \nabla\cdot\mathbf{J}_0 + C_1^{\nabla J} \rho_1 \nabla\cdot\mathbf{J}_1, \qquad (3)$$

where

$$\mathbb{J}^2 = \sum_{\rho\sigma} J_{\rho\sigma}^2, \text{ and} \qquad (4)$$

$$J_{\rho\sigma} = \frac{1}{3} J^{(0)} \delta_{\rho\sigma} + \frac{1}{2}\boldsymbol{\varepsilon_{\rho\sigma\tau}} J_\tau + J_{\rho\sigma}^{(2)} \qquad (5)$$

If we impose spherical symmetry, the scalar part $J^{(0)}$ and symmetric-tensor part $J_{\rho\sigma}^{(2)}$ of spin-current density vanish and we are left with

$$H_T = \frac{1}{2} C_0^J J_0^2(r) + \frac{1}{2} C_1^J J_1^2(r) \qquad (6)$$

$$H_{SO} = -C_0^{\nabla J} J_0(r)\frac{d\rho_0}{dr} - C_1^{\nabla J} J_1(r)\frac{d\rho_1}{dr} \qquad (7)$$

After variation and manipulation one obtains the expressions for proton (- neutron) spin-orbit potential as

$$V_{SO}^q = \frac{1}{2r}\left\{(C_0^J - C_1^J)J_0(r) + 2C_1^J J_q(r) - (C_0^{\nabla J} - C_1^{\nabla J})\frac{d\rho_0}{dr} - 2C_1^{\nabla J}\frac{d\rho_q}{dr}\right\}\mathbf{L}\cdot\mathbf{S} \qquad (8)$$

Following Lesinsky et al [20] the structure of the tensor part of the EDF can be recast as

$$H_T = \frac{1}{2}\alpha(\mathbf{J}_n{}^2 + \mathbf{J}_p{}^2) + \beta \mathbf{J}_n \cdot \mathbf{J}_p \tag{9}$$

with

$$\alpha = C_0^J + C_1^J, \quad \beta = C_0^J - C_1^J \tag{10}$$

and omitting the iso-vector part, the spin-orbit potential component is given by

$$V_{s.o.}^q = \frac{W_0}{2r}\left(2\frac{d\rho_q}{dr} + \frac{d\rho_{q'}}{dr}\right) + \left(\alpha \frac{J_q}{r} + \beta \frac{J_{q'}}{r}\right) \tag{11}$$

where $J_{q(q')}(r)$ is the proton or neutron spin-orbit density defined as

$$J_{q(q')}(r) = \frac{1}{4\pi r^3}\sum_i v_i^2 (2j+1)\left[j_i(j_i+1) - l_i(l_i+1) - \frac{3}{4}\right] R_i^2(r) \tag{12}$$

In this expression q stands for neutrons(protons) and q' for protons(neutrons), and i = n,l,j runs over all states having the given q(q') and $R_i(r)$ is the radial part of the wave function. $v_i^2$ is the BCS occupation probability of each orbital and

$$C_0^{\nabla J} = -\frac{3}{4} W_0. \tag{13}$$

The terms in the first part of eq.(11) come directly from the Skyrme spin-orbit interaction, and the terms in the second parentheses include both the central exchange and the tensor contributions. There are certain features associated with the tensor and the central exchange contributions to the spin-orbit splitting. Firstly, the A-dependence or isospin dependence of the first and second term in the r.h.s. of eq. (1) needs to be mentioned. Because of the contribution from the Skyrme spin-orbit potential (proportional to $W_0$) is linear in the first order derivatives of the proton and neutron density distributions the associated mass number and isospin dependence is very moderate for heavy nuclei. On the other hand spin-orbit current $J_q$ shows shell effect giving strong fluctuations. $J_q$ gives essentially no contribution for the spin-saturated cases where all spin-orbit partners are either completely occupied or empty. But the contribution from $J_q$ increases linearly with the number of particles if only one of the spin-orbit partners $J_>$ or $J_<$ is filled.

In eq.(11) $\alpha = \alpha_c + \alpha_T$ and $\beta = \beta_c + \beta_T$. The central exchange contributions are written in terms of the usual Skyrme force parameters as

$$\alpha_c = \frac{1}{8}(t_1 - t_2) - \frac{1}{8}(t_1 x_1 + t_2 x_2) \tag{14}$$

$$\beta_c = -\frac{1}{8}(t_1 x_1 + t_2 x_2) \tag{15}$$

The tensor contributions are expressed as

$\alpha_T = \frac{5}{12} U$, $\beta_T = \frac{5}{24}(T + U)$, where the coupling constants T and U denote the strength of the triplet-even and triplet-odd tensor interactions respectively.

To observe the effect of tensor force on the ground state properties of nuclei across the nuclear chart through Skyrme like mean-field calculation what is usually done is that several constraints like binding energies, r.m.s. charge radii, nuclear matter density, compressibility modulus etc. are used in getting the best fit force parameters. However, the effect of the tensor interaction has been included in a perturbative approach by several authors where all other parameters are kept in a 'frozen' condition. According to Stancu et al [4] the optimal parameters of tensor interaction, viz., $\alpha_T$ and $\beta_T$ should be found from a triangle in the two dimensional ($\alpha_T$, $\beta_T$) plane, lying in the quadrant of negative $\alpha_T$ and positive $\beta_T$. Colo et al [17] have refitted the values of ($\alpha_T$, $\beta_T$) using the experimental data [29] for the single particle states in the N = 82 isotones and the Z = 50 isotopes. The values are $\alpha_T$ = -170 Mev fm$^5$ and $\beta_T$ = 100 MeV fm$^5$. They did not try to refit all the Skyrme parameters after including the tensor term. From another perturabative study of Brink and Stancu [18] where they have used SIII parameters for Skyrme forces and varied $\alpha$ and $\beta$ only keeping other force parameters unaltered, we get $\alpha_T$ = -180 MeV fm$^5$ and $\beta_T$ = 120 MeV fm$^5$. Moreno-Torres et al [22] have also used the prescription of Colo et al [17] in the calculation for proton/ neutron shell gap for Z,N = 8, 20, 28.

We have used SKP set of parameters for the whole set of our calculations. The reasons behind the use of SKP set are: i) it takes the $\mathbf{J}^2$ terms from the central force into account which is necessary for inclusion of the tensor force, ii) it reproduces the ground state properties and the single particle structure of $^{208}$Pb, the heaviest stable doubly magic nucleus, both theoretically and experimentally one of the most well studied nuclei, in a way much better than any other parameter set as can be seen later. The utility of SKP parameter set was also highlighted by Zalewski et al [21] as it contains the usual $C_t^J$ – coupling constant terms in the energy density functional (EDF), though they have mentioned that their findings were independent of the variants of Skyrme force parameters. Another point is to be noted here, the detailed calculations [21] have shown the impact of polarization effects on the spin-orbit splitting is indeed very small reflecting a cancellation of polarization effects exerted on the j = l ± ½ which guided us to calculate the spin-orbit splitting of shell closed nuclei without considering the mass (time-even), shape (time-even) and spin (time-odd) polarization effects. A schematic pairing force has been used through the energy functional

$$E_{pair} = \sum_q G_q \left[ \sum_{\beta \in q} \sqrt{w_\beta (1 - w_\beta)} \right]^2 ,$$

(16)

Where the pairing matrix elements $G_q$ are constant within each species q ∈ {π, ν} and $w_\beta$ are the occupation probabilities of the shell model states.

In order to evaluate the effect of the tensor interaction on evolution of shells we have followed the prescription of Zalewski et al by calculating the spin-orbit splitting of several shell closed nuclei. To get the optimized coupling constant $C_0^{\nabla J}$, variational studies of the spin-orbit splitting of $1f_{7/2} - 1f_{5/2}$, $2p_{3/2} - 2p_{1/2}$ and $1d_{5/2} - 1d_{3/2}$ levels were performed for isoscalar spin-saturated N = Z nucleus $^{40}$Ca. It has been observed that the separation energies of the spin-orbit doublets decrease with reduction in values of $W_0$ (-4/3 $C_0^{\nabla J}$ ) and a reduction ~ 20% is required for agreement with the experimental results for neutron Δ1f. After fixing the first coupling constant $C_0^{\nabla J}$ we moved to $^{56}$Ni which is a spin-unsaturated isoscalar nucleus to get the optimized isoscalar tensor coupling constant $C_0^J$. Reasonably good fit to spin-orbit splitting of 1f and 2p levels were obtained (Fig.1). Then the isovector tensor coupling constant $C_1^J$ was fixed by reproducing the spin-orbit level splittings of 1f, 2p and 1d levels of N ≠ Z nucleus $^{48}$Ca. One comment is pertinent here, if we critically survey the information compiled by Zalewski et al in their Table III one can find easily that the values of spin-orbit splitting of neutron and proton states of same $l$ values are almost identical for all the doubly shell closed nuclei. Keeping in mind the fact just stated it seems that the value of difference in energy of 1f neutron states is a bit different compared to energy of proton 1f doublet in quoted values of $^{48}$Ca. Our final adjustment of the tensor coupling constants was done after calculating the spin-orbit splitting of several shell model states near the Fermi-surface of the stable shell closed nuclei $^{208}$Pb. As $^{208}$Pb is experimentally a very well studied nucleus, the energies of shell model doublets are more reliable compared to single particle energies in other nuclei.

### III. RESULTS AND DISCUSSION:

In Table I we present the scenario of single particle spectrum for $^{208}$Pb calculated under different parameter sets. The single particle behaviour of these states has been well established from the spectroscopic measurements [30, 31]. It is quite apparent from the table that SKP produces the best single particle (-hole) spectrum both for proton as well as neutron states of $^{208}$Pb. Out of eighteen states studied only in the cases of proton $1g_{9/2}$ state and neutron $1h_{9/2}$ and $1i_{11/2}$ states we find some discrepancies of the order of 1 MeV, otherwise there is a very good agreement.

One of the most important ground state properties of any nucleus is its charge density distribution. The charge density distribution of $^{208}$Pb has long been investigated by several experimental groups using different methods. Furthermore, the charge distribution of $^{208}$Pb measured through electron scattering and muonic X-ray method had been subjected to model independent Fourier-Bessel analysis,

$\rho(r) = \sum_\mu a_\mu j_0(q_\mu r)$, for $r \leq R$

$\phantom{\rho(r)} = 0$, for $r > R$ (17).

Here $j_0(q_\mu r)$ are the spherical Bessel functions. In Table II we present our Fourier-Bessel analysis of the calculated charge distribution of $^{208}$Pb. A close correspondence with the experimental values [32] has been observed for first nine coefficients. The r.m.s. charge radius for $^{208}$Pb has been obtained from the calculated charge distribution. The value came out to be 5.498 fm. whereas the experimental value is 5.501 fm.

The variation of splitting of the neutron and proton spin-orbit doublet $1f_{7/2} - 1f_{5/2}$ of $^{40}$Ca with $W_0$, the central spin-orbit strength, which is related to tensor coupling constant $C_0^{\nabla J}$ (eq.13) has been presented in Table III from where we find that the agreement between empirical [33,34] and theoretical values for neutron states improves as $W_0$ is decreased. A similar study in the cases for $2p$ and $1d$ doublets has been made (Table IV). But here we find the worsening of agreement with lower values of $W_0$. Keeping in view of the reproduction of the spin-orbit splitting of some shell model states of doubly shell closed nuclei in the whole nuclear chart in general and of $^{208}$Pb in particular, optimization of the three tensor parameters has been done. All other force parameters were kept at their Skyrme level. The difference in energies for some shell model states near the Fermi-surface of $^{208}$Pb are shown in Table V for a few standard Skyrme force parameters from where we see that after inclusion of tensor interaction in SKP set, a good agreement with empirical values has been achieved. In order to verify the reliability of our method of using SKP-T we present in Table VI the calculated splitting of the spin-orbit partners of proton and neutron single particle states $^{16}$O, $^{40,48}$Ca, $^{56}$Ni, $^{100,132}$Sn along with the empirical values [33, 34].

For the doubly shell closed nucleus $^{132}$Sn the gap observed between the spin-orbit doublet $(\nu)2f$ is 2.58 MeV while the experimental value is 2.01 MeV. Considering the uncertainties involved in the experimental determination of single particle energies to be ~ 15 – 20%, our calculated values are not that far off. It has been observed in the works of Brink and Stancu [18] and also of Colo et al [17] that for the energy difference of the single particle $1h_{11/2}$ and $1g_{7/2}$ states of Sn – isotopes the effect of tensor force is to increase the gap with increasing neutron number whereas there is a flat pattern for Skyrme forces without any tensor component. Fig. 2 shows the results with SKP parameters with and without tensor terms for the evolution of the energy difference for Sn-isotopes. It is clear from the figure that with the inclusion of tensor interaction we get the upward trend of the gap evolution but the result is not that dramatic as observed in the case of experimental values [35]. From N = 64 to 72 there are some differences between the calculated values and experimental values but the situation improves after N = 72. For the doubly shell closed nucleus $^{132}$Sn there is no contribution from the tensor part in the spin-orbit splitting as in this case $J_p \cong J_n$.

In Fig. 3 we present the differences in energies of the $1d_{5/2}$ and $2s_{1/2}$ shell model states with respect to $1d_{3/2}$ state for calcium isotopes (40 – 48) with and without the tensor force. It is evident that the inclusion of tensor interaction improves the splitting of the spin-orbit partners considerably showing the correct pattern and the behaviour of calculated $2s_{1/2}$ states reflects also the trend observed in experiments [36]. Ordering of single particle states is also faithfully preserved.

Neutron single-particle energy differences between $1i_{13/2}$ and $1h_{9/2}$ states in N = 82 isotones calculated with and without inclusion of tensor force along with experimental values [35] have been presented in Fig. 4. Calculation without the tensor force shows the opposite behaviour to that of the experimental graph. The importance of the tensor interaction can be visualized from the figure. The splitting of the spin-orbit partners $1h_{11/2}$ and $1h_{9/2}$ states is very important for the development of the N = 82 shell and in case of Tin isotopes the tensor interaction produces the correct spin-orbit potential which generates the shell gap.

In Fig. 5 we present the gap evolution for Z = 8 with and without tensor forces. For comparison experimental data sets have also been shown. The gap here means the difference in single particle energies of the first particle state above and the last hole state below the Fermi surface of the nuclei under consideration. Since we are dealing with shell closed nuclei, the shell model nomenclature has been used for the particle (-hole) states. For the Z = 8 isotopes and N = 8 isotones the energy difference between the first unoccupied proton (neutron) level $1d_{5/2}$ and the last occupied state $1p_{1/2}$ constitute the gap. Inclusion of tensor interaction changes the curvature of the shell evolution of the graph correctly and also produces the kink at N = 16 indicating a shell closure at $^{24}$O as observed in the recent experimental results.

Evolution of the gap between the last occupied $1p_{1/2}$ and first level above the Fermi-surface $1d_{5/2}$ neutron levels for Z = 6 to Z = 12 (N = 8 isotones) has been presented in Fig. 6. Except for the sharp kink in gap energy at shell closure Z = 8 the calculation with tensor interaction produces a fair representation of the experimental situation.

The gap for Z = 20 isotopes and N = 20 isotonic chain is the energy difference between the proton (neutron) single particle states $1f_{7/2}$ and $1d_{3/2}$. In the gap evolution of Z = 20 isotopes as shown in Fig. 7, effect of tensor interaction increases the gap as more and more neutrons are added in the system. We find the gap reaches a maximum at N = 32. Dinca et al [37] have studied the even $^{52-56}$Ti isotopes with intermediate-energy Coulomb excitation and absolute $B(E2; 0^+ \to 2_1^+)$ transition rates have been obtained. These data confirm the presence of a sub-shell closure at neutron number $N = 32$ in neutron-rich nuclei above the doubly magic nucleus $^{48}$Ca.

In fig. 8 the evolution of gaps in $N = 20$ isotones has been presented. Though in the experimental results [38] one can observe ups and downs we get a slow increment up to $Z = 20$ and then a faster increment for $Z > 20$ isotones whereas without inclusion of tensor interaction there is almost a flat behaviour of the gap evolution. May be some more physics input are needed.

One interesting feature of our mean-field calculation Table VII shows the comparative study of the splitting of fully occupied shell model states below the Fermi - surface in $^{40,48}$Ca, $^{68,78}$Ni, $^{90,110}$Zr, $^{120,132}$Sn, $^{208}$Pb. In these nuclei, either proton and neutron shells are both spin saturated or one of them is saturated while the other is unsaturated, or else both of them are unsaturated. When both proton and neutron shells are spin-saturated, there is no effect of the tensor force as $J_p$ and $J_n$ do not contribute in the spin-orbit potential. As a result of that the spin-obit splitting for fully occupied states remain unaltered as can be seen in the case of $^{40}$Ca and $^{110}$Zr. Even other ground state properties like $R_c$, $R_n$ do not change. When either of proton and neutron shells is spin-saturated, tensor contribution in the spin-orbit potential becomes substantial and as a result of that there is a marked change in the level splitting of the spin-orbit partners as can be observed in the cases of $^{48}$Ca, $^{68}$Ni, $^{90}$Zr, $^{120}$Sn. However, the most interesting observation is that when both proton and neutron shells are spin – unsaturated there is almost no effect of the tensor interaction in the splitting of the spin-orbit partners for the nuclei $^{78}$Ni, $^{132}$Sn and $^{208}$Pb. So, these nuclei are somewhat special in the club of doubly shell closed nuclei. The reason behind this observation is the presence of spin-orbit densities $J_p$ and $J_n$ in the spin-orbit potential $V_{so}$ (eq. 11). For the aforementioned nuclei calculated values of $J_p$ and $J_n$ using SKP are both positive with comparable magnitudes and $\alpha \cong -\beta$. There is another point to be noted : energy differences of spin-orbit partners of the same shell model states of protons and neutrons for both spin-saturated and both spin-unsaturated cases are almost similar. This feature is also evident in the Table III of Zalewski et al [21]. From figures 9 and 10 where $J_p$ and $J_n$ have been drawn for the nuclei discussed earlier it can be seen that for $N = Z$ nuclei, $J_p$ and $J_n$ are almost equal and as a result of that there is no effect of tensor interaction in the spin-orbit splitting of shell model states in these nuclei.

## IV. CONCLUSION

In this paper the importance of the tensor interaction on the spin-orbit splitting of shell model states leading to evolution of shell gaps in different region of the nuclear chart have been investigated using SKP and SKP + T parameters. Reproduction of the observed splittings of spin-orbit partners for the single particle(-hole) states of $^{208}$Pb along with $^{40,48}$Ca and $^{56}$Ni has been utilized to get the suitable tensor coefficients. It has been observed that the optimized coefficients of tensor interaction reproduces the evolution of shell gaps for Z,N = 8, 20 and energy gaps between $1h_{11/2}$ and $1g_{7/2}$ states of Sn-isotopes, energy differences between $1i_{13/2}$ and $1h_{11/2}$ states of N = 82 isotones, energy differences between $1d_{5/2}$, $2s_{1/2}$ and $1d_{3/2}$ states of Ca isotopes as observed in the works of Wang et al [24] and Brink and Stancu [18] which reiterates our stand that for evolution of shells, reproduction of splitting of shell model states of $^{208}$Pb is a good starting point.

Furthermore, the spin-orbit splittings of fully occupied states of doubly shell closed nuclei $^{20,28}$Ca, $^{68,78}$Ni, $^{90,110}$Zr, $^{120,132}$Sn and $^{208}$Pb have been calculated which reveal that for both spin-saturated proton and neutron shell closed nuclei and for both spin-unsaturated proton and neutron shell closed nuclei, the splitting of the shell model states are almost similar with or without inclusion of tensor interaction. However, the situation changes when either one of the proton or neutron shells is spin-saturated while the other one is not.

ACKNOWLEDGEMENT        The author thanks the University Grants Commission for support by the Emeritus Fellowship [ No. F.6-34/2011(SA – II)].

TABLE I

Single particle levels in $^{208}$Pb

| nlj | $-E_{nlj}$ (MeV) | | | | | |
|---|---|---|---|---|---|---|
| | SKM | $Z_\sigma$ | Sly4 | SKX | SKP | EXPT |
| *Proton* | | | | | | |
| $1g_{9/2}$ | 16.12 | 17.50 | 16.44 | 16.15 | 15.18 | 16.03 |
| $1g_{7/2}$ | 12.32 | 13.34 | 14.67 | 11.36 | 11.43 | 11.51 |
| $2d_{5/2}$ | 10.16 | 10.88 | 11.25 | 9.64 | 10.33 | 10.23 |
| $2d_{3/2}$ | 8.28 | 8.87 | 7.07 | 7.54 | 8.80 | 8.38 |
| $3s_{1/2}$ | 7.56 | 8.04 | 9.01 | 7.04 | 8.11 | 8.03 |
| $1h_{11/2}$ | 8.42 | 9.58 | 8.11 | 9.95 | 8.72 | 9.37 |
| $1h_{9/2}$ | 2.95 | 3.62 | 5.54 | 3.07 | 3.52 | 3.60 |
| $2f_{7/2}$ | 1.81 | 2.13 | 2.32 | 2.47 | 3.21 | 2.91 |
| *Neutron* | | | | | | |
| $1h_{11/2}$ | 17.27 | 18.09 | 16.32 | 16.52 | 15.03 | 14.50 |
| $1h_{9/2}$ | 11.73 | 12.09 | 14.09 | 9.46 | 9.84 | 11.28 |
| $2f_{7/2}$ | 11.50 | 11.68 | 11.21 | 9.98 | 10.50 | 10.38 |
| $2f_{5/2}$ | 8.53 | 8.40 | 9.81 | 6.67 | 8.11 | 7.95 |
| $3p_{3/2}$ | 8.56 | 8.63 | 8.65 | 7.10 | 8.21 | 8.27 |
| $3p_{1/2}$ | 7.40 | 7.33 | 8.10 | 5.81 | 7.32 | 7.38 |
| $1i_{13/2}$ | 9.42 | 9.55 | 7.06 | 10.37 | 8.51 | 9.38 |
| $1i_{11/2}$ | 2.07 | 1.83 | 3.82 | 1.02 | 1.84 | 3.15 |
| $2g_{9/2}$ | 3.35 | 2.92 | 1.64 | 3.19 | 3.71 | 3.74 |
| $2g_{7/2}$ | 0.27 | 1.03 | 0.04 | 0.95 | 0.71 | 1.45 |

## TABLE II

Fourier-Bessel coefficients of charge distribution of $^{208}$Pb

| ν | $a_\nu \times (10^{-3})$ | |
|---|---|---|
| | SKP | EXPT |
| 1 | 0.64176 | 0.633310 |
| 2 | 0.62155 | 0.61907 |
| 3 | -0.47033 | -0.48344 |
| 4 | -0.40295 | -0.34409 |
| 5 | 0.48581 | 0.35260 |
| 6 | 0.021915 | 0.12061 |
| 7 | -0.17691 | -0.17545 |
| 8 | -0.04983 | -0.11982 |
| 9 | 0.25197 | 0.88501 |

Table III

Variation of splitting of $^{40}$Ca 1f shell model states with central spin-orbit strength

| $W_0$ | Neutron | Proton |
|---|---|---|
|  | $\Delta$1f | $\Delta$1f |
| 80.0 | 5.17 | 4.91 |
| 90.0 | 5.85 | 5.56 |
| 100.0 | 6.56 | 6.23 |
| Expt. | 5.64 | 6.05 |

Table IV

Variation of splitting of $^{40}$Ca 2p and 1d shell model states with central spin-orbit strength

| $W_0$ | Neutron | | Proton | |
|---|---|---|---|---|
|  | $\Delta$2p | $\Delta$1d | $\Delta$2p | $\Delta$1d |
| 80.0 | 1.49 | 4.10 | 1.35 | 3.98 |
| 90.0 | 1.69 | 4.66 | 1.52 | 4.52 |
| 100.0 | 1.89 | 5.24 | 1.70 | 5.10 |
| Expt. | 2.00 | 6.00 | 1.69 | 5.94 |

TABLE V

Splitting of spin-orbit doublets in $^{208}$Pb

| Protons | Energy in MeV | | | | | | |
|---|---|---|---|---|---|---|---|
| | Sly5 | Sly4 | SKM$^*$ | SIII | SKP | SKP-T | EXP |
| Δ1h | 6.43 | 6.22 | 5.94 | 5.20 | 5.24 | 5.30 | 5.56 |
| Δ2d | 1.95 | 1.89 | 2.37 | 1.63 | 1.55 | 1.41 | 1.33 |
| Δ2f | 2.69 | 2.61 | 2.57 | 2.32 | 2.15 | 1.98 | 1.93 |
| Δ3p | 1.05 | 1.02 | 0.97 | 0.87 | 0.76 | 0.69 | 0.84 |
| Neutrons | | | | | | | |
| Δ1h | 5.83 | 5.59 | 5.55 | 4.73 | 5.18 | 5.09 | 5.10 |
| Δ1i | 7.65 | 7.25 | 7.26 | 6.39 | 6.62 | 6.39 | 6.46 |
| Δ2f | 2.05 | 1.96 | 2.93 | 2.67 | 2.30 | 2.12 | 2.03 |
| Δ2g | 3.68 | 3.57 | 3.58 | 3.30 | 2.93 | 2.73 | 2.51 |
| Δ3p | 1.17 | 1.13 | 1.13 | 1.02 | 0.83 | 0.74 | 0.90 |
| Δ3d | 1.72 | 1.67 | 1.60 | 1.47 | 1.27 | 1.15 | 0.97 |

Table VI

Calculated values of spin-orbit splitting along with empirical values of doubly shell closed nuclei O, Ca, Sn

| Nucleus | Orbitals | Energy difference(T) | Energy difference(E) |
|---|---|---|---|
| $^{16}$O | νΔ1p | 4.18 | 6.17 |
|  | νΔ1d | 4.89 | 5.08 |
|  | πΔ1p | 4.06 | 6.32 |
|  | πΔ1d | 4.65 | 4.97 |
| $^{40}$Ca | νΔ2p | 1.87 | 2.00 |
|  | νΔ1f | 6.45 | 5.64 |
|  | νΔ1d | 5.13 | 6.00 |
|  | πΔ2p | 1.65 | 1.72 |
|  | πΔ1f | 6.13 | 6.05 |
|  | πΔ1d | 4.97 | 5.94 |
| $^{48}$Ca | νΔ2p | 2.33 | 1.77 |
|  | νΔ1f | 7.59 | 8.01 |
|  | νΔ1d | 6.05 | 5.30 |
|  | πΔ2p | 1.24 | 2.14 |
|  | πΔ1f | 5.03 | 4.92 |
|  | νΔ1d | 3.49 | 5.29 |
| $^{56}$Ni | νΔ2p | 1.83 | 1.88 |
|  | νΔ1f | 6.42 | 7.16 |
|  | πΔ2p | 1.58 | 1.83 |
|  | νΔ1f | 6.09 | 7.01 |
| $^{90}$Zr | νΔ2d | 2.65 | 2.43 |
|  | νΔ1g | 7.34 | 7.07 |
|  | νΔ2p | 1.77 | 0.37 |
|  | ν∆1f | 5.71 | 1.71 |
|  | π∆2d | 1.69 | 2.03 |
|  | π∆1g | 5.28 | 5.56 |
|  | π∆2p | 1.21 | 1.50 |
|  | π∆1f | 3.84 | 4.56 |
| $^{100}$Sn | ν∆2d | 2.09 | 1.93 |
|  | ν∆1g | 6.39 | 7.00 |
|  | π∆1g | 5.98 | 6.82 |
|  | π∆2p | 1.26 | 2.85 |
| $^{132}$Sn | ν∆2f | 2.68 | 2.00 |
|  | ν∆3p | 0.93 | 0.81 |
|  | ν∆1h | 7.37 | 6.68 |
|  | ν∆2d | 2.08 | 1.93 |
|  | π∆2d | 1.72 | 1.75 |
|  | π∆1g | 4.62 | 5.33 |

Empirical values are taken from ref. [33] and [34]

TABLE VII

Ground state properties of some shell closed nuclei including spin-orbit splitting of completely filled up states below the Fermi surface

| Nucleus | SKP | | | | SKP-T | | | |
|---|---|---|---|---|---|---|---|---|
| | $R_c$ (fm) | $R_n$ (fm) | $\Delta E$ (p) (MeV) | $\Delta E$ (n) (MeV) | $R_c$ (fm) | $R_n$ (fm) | $\Delta E$(p) (MeV) | $\Delta E$(n) (MeV) |
| $^{40}$Ca (p-s/n-s) | 3.54 | 3.40 | $\Delta 1d$ = 4.98 | $\Delta 1d$ = 5.13 | 3.54 | 3.40 | $\Delta 1d$ = 4.99 | $\Delta 1d$ = 5.12 |
| $^{48}$Ca (p-s/n-u) | 3.55 | 3.64 | $\Delta 1d$ = 5.57 | $\Delta 1d$ = 4.09 | 3.54 | 3.61 | $\Delta 1d$ = 3.10 | $\Delta 1d$ = 6.50 |
| $^{68}$Ni (p-u/n-s) | 3.95 | 4.03 | $\Delta 1d$ = 3.20 | $\Delta 1d$ = 4.70<br>$\Delta 1f$ = 6.67<br>$\Delta 2p$ = 1.80 | 3.94 | 4.02 | $\Delta 1d$ = 4.94 | $\Delta 1d$ = 2.77<br>$\Delta 1f$ = 4.52<br>$\Delta 2p$ = 1.16 |
| $^{78}$Ni (p-u/n-u) | 4.00 | 4.21 | $\Delta 1d$ = 3.41 | $\Delta 1d$ = 3.39<br>$\Delta 1f$ = 5.49<br>$\Delta 2p$ = 1.47 | 4.00 | 4.21 | $\Delta 1d$ = 3.19 | $\Delta 1d$ = 3.15<br>$\Delta 1f$ = 5.66<br>$\Delta 2p$ = 1.24 |
| $^{90}$Zr (p-s/n-u) | 4.31 | 4.30 | $\Delta 2p$ = 1.54<br>$\Delta 1f$ = 5.63 | $\Delta 2p$ = 1.43<br>$\Delta 1f$ = 4.24 | 4.30 | 4.29 | $\Delta 2p$ = 1.16<br>$\Delta 1f$ = 3.33 | $\Delta 2p$ = 1.83<br>$\Delta 1f$ = 6.45 |
| $^{110}$Zr (p-s/n-s) | 4.47 | 4.71 | $\Delta 2p$ = 1.38<br>$\Delta 1f$ = 4.10 | $\Delta 2p$ = 1.56<br>$\Delta 1f$ = 2.32 | 4.47 | 4.71 | $\Delta 2p$ = 1.40<br>$\Delta 1f$ = 4.03 | $\Delta 2p$ = 1.55<br>$\Delta 1f$ = 2.27 |
| $^{120}$Sn (p-u/n-s) | 4.69 | 4.73 | $\Delta 2p$ = 1.17<br>$\Delta 1f$ = 3.24 | $\Delta 2p$ = 1.49<br>$\Delta 2d$ = 2.37<br>$\Delta 1f$ = 4.79<br>$\Delta 1g$ = 6.48 | 4.68 | 4.72 | $\Delta 2p$ = 1.56<br>$\Delta 1f$ = 4.98 | $\Delta 2p$ = 1.12<br>$\Delta 2d$ = 1.67<br>$\Delta 1f$ = 2.98<br>$\Delta 1g$ = 4.44 |
| $^{132}$Sn (p-u/n-u) | 4.73 | 4.88 | $\Delta 2p$ = 1.13<br>$\Delta 1f$ = 3.69 | $\Delta 2p$ = 1.25<br>$\Delta 2d$ = 2.00<br>$\Delta 1f$ = 3.69<br>$\Delta 1g$ = 5.43 | 4.73 | 4.88 | $\Delta 2p$ = 1.21<br>$\Delta 1f$ = 3.59 | $\Delta 2p$ = 1.14<br>$\Delta 2d$ = 1.67[a]<br>$\Delta 1f$ = 3.55<br>$\Delta 1g$ = 5.57 |
| $^{208}$Pb (p-u/n-u) | 5.497 | 5.554 | $\Delta 2p$ = 0.88<br>$\Delta 2d$ = 1.52<br>$\Delta 1f$ = 2.41<br>$\Delta 1g$ = 3.74 | $\Delta 3p$ = 0.90<br>$\Delta 2p$ = 0.98<br>$\Delta 2d$ = 1.73<br>$\Delta 2f$ = 2.41<br>$\Delta 1g$ = 3.74<br>$\Delta 1h$ = 5.19 | 5.497 | 5.554 | $\Delta 2p$ = 0.93<br>$\Delta 2d$ =1.63[c]<br>$\Delta 1f$ = 2.30<br>$\Delta 1g$ = 3.41 | $\Delta 3p$ = 0.91[b]<br>$\Delta 2p$ = 0.80<br>$\Delta 2d$ = 1.59<br>$\Delta 2f$ = 2.19[d]<br>$\Delta 1g$ = 3.71<br>$\Delta 1h$ = 5.09[e] |

p-s = proton spin-saturated, p-u = proton spin-unsaturated; n-s = neutron spin-saturated, n-u = neutron spin-unsaturated

Experimental values – a) $\Delta 2d$ = 1.75 [39]

b) $\Delta 3p$ = 0.90, c) $\Delta 2d$ = 1.33, d) $\Delta 2f$ = 2.13, e) $\Delta 1h$ = 5.10 [29]

# FIGURE CAPTION

Fig. 1. "(Color online) Spin orbit splittings of $1f$ and $2p$ states of $^{40}$Ca.

Fig. 2. "(Color online)" Energy differences of $1h_{11/2}$ and $1g_{9/2}$ states of Sn-isotopes.

Fig. 3. "(Color online)" Energy splitting of $1d_{5/2}$ and $2s_{1/2}$ states of Ca-isotopes with respect to $1d_{3/2}$ state.

Fig. 4. "(Color online)" Energy differences of $1i_{13/2}$ and $1h_{9/2}$ states of N = 82 isotones.

Fig. 5. "(Color online)" Comparison of calculated gap evolution of Z = 8 isotones with and without tensor interaction and experimental data.

Fig. 6. "(Color online)" Comparison of calculated gap evolution of N = 8 isotopes with and without 6ensor interaction and experimental data.

Fig. 7. "(Color online)" Comparison of calculated gap evolution of Z = 20 isotones with and without tensor interaction and experimental data.

Fig. 8. "(Color online)" Comparison of calculated gap evolution of N = 20 isotones with and without tensor interaction and experimental data.

Fig. 9. "(Color online)" Spin-orbit density for proton states of $^{16}$O, $^{40,48}$Ca, $^{56}$Ni, $^{100,132}$Sn and $^{208}$Pb.

Fig. 10. "(Color online)" Spin-orbit density for neutron states of $^{16}$O, $^{40,48}$Ca, $^{56}$Ni, $^{100,132}$Sn and $^{208}$Pb.

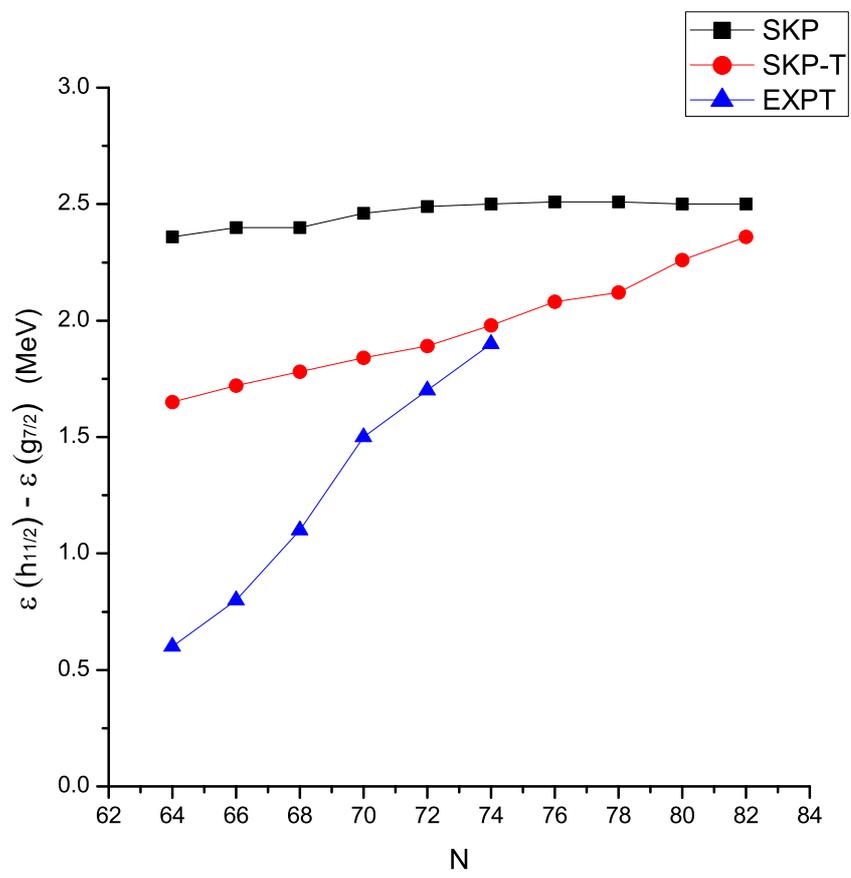

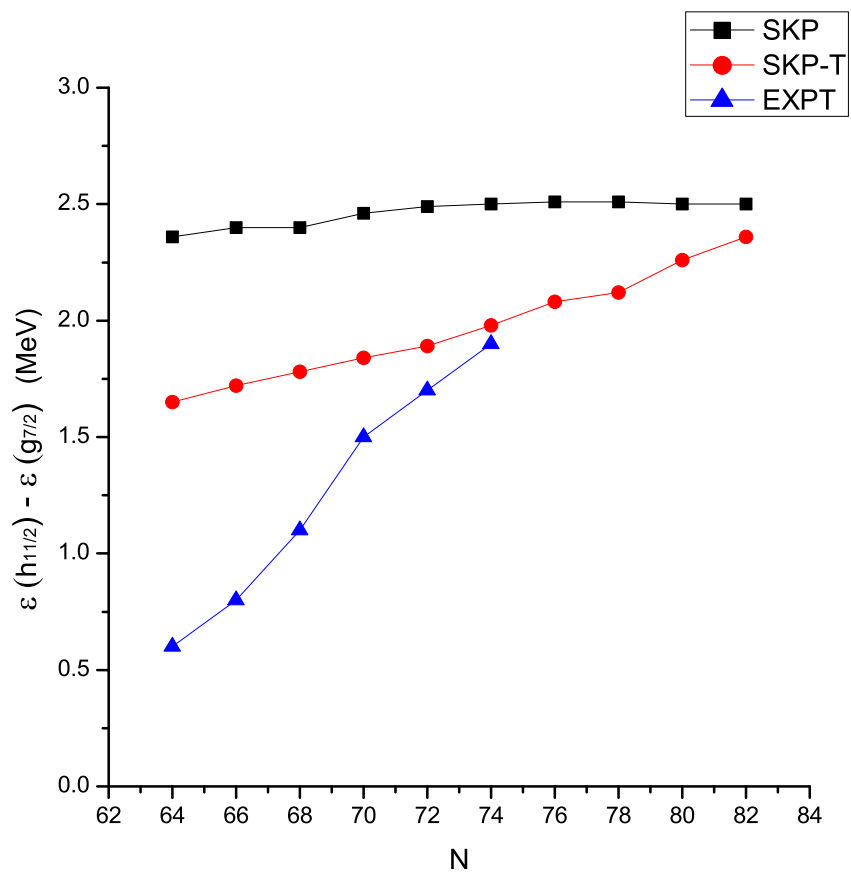

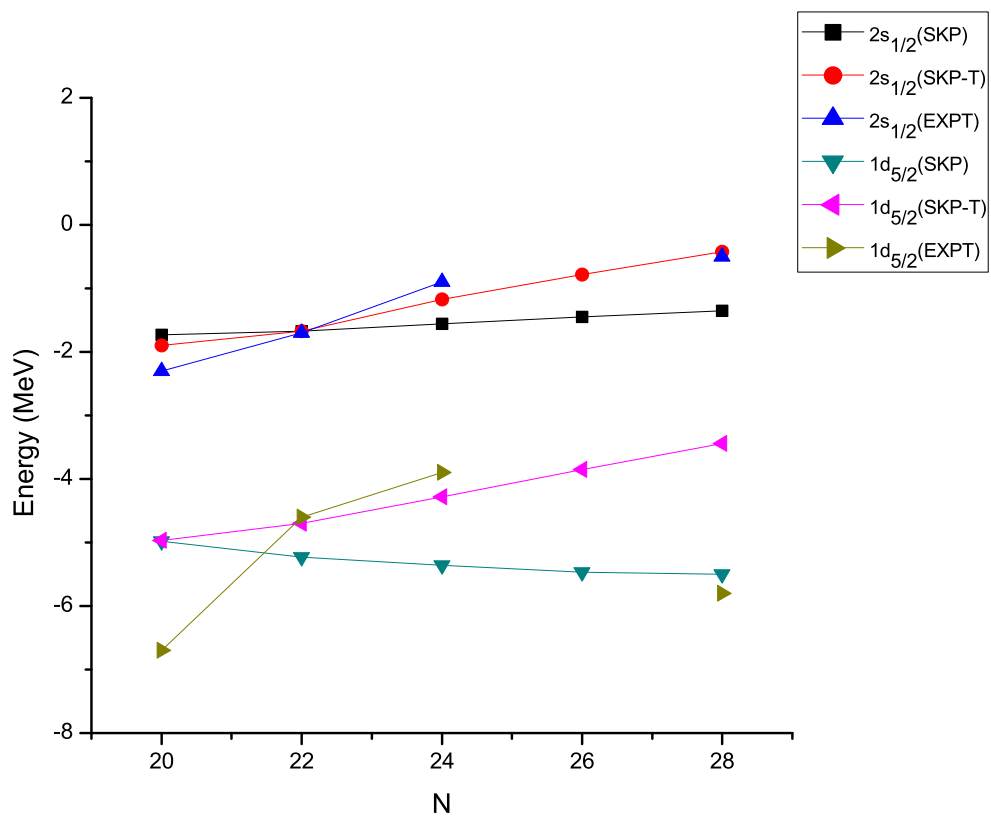

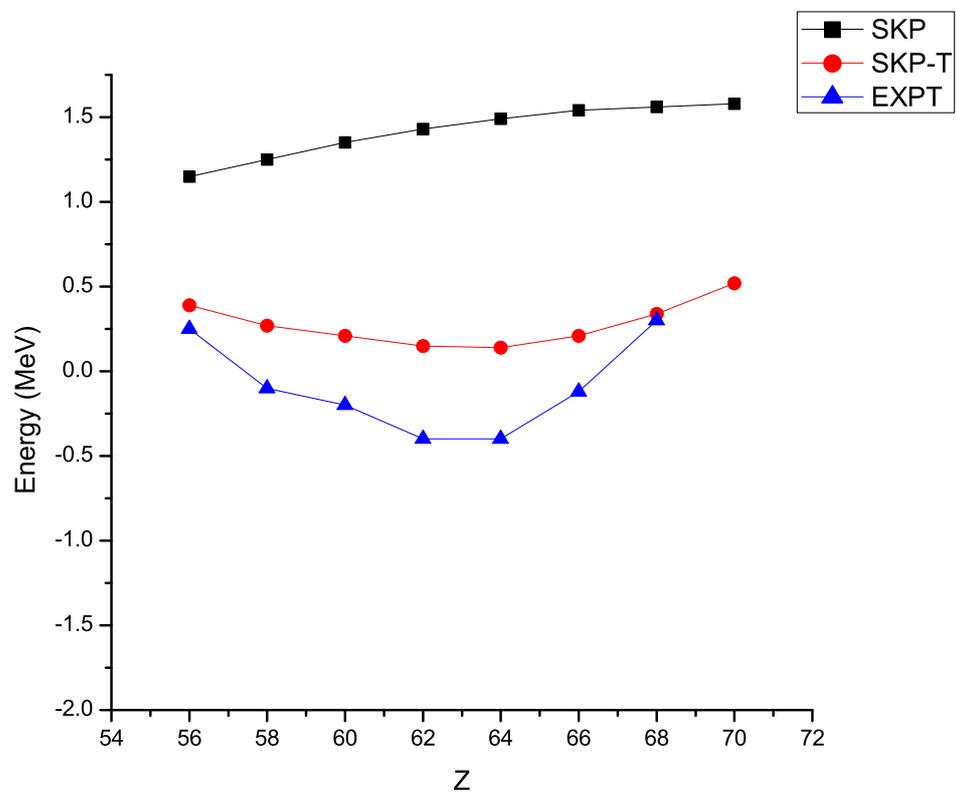

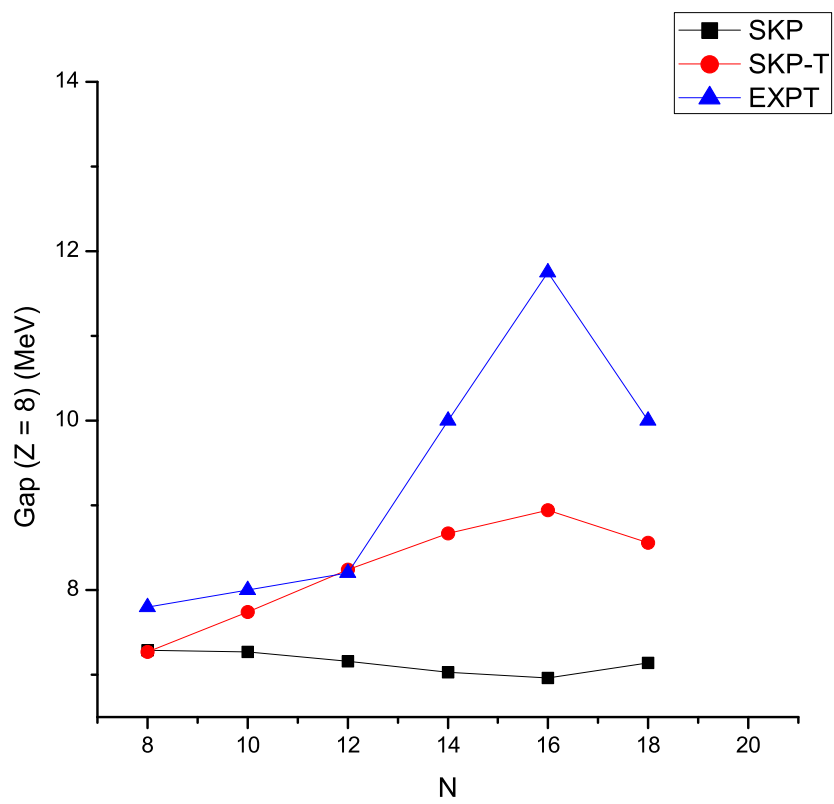

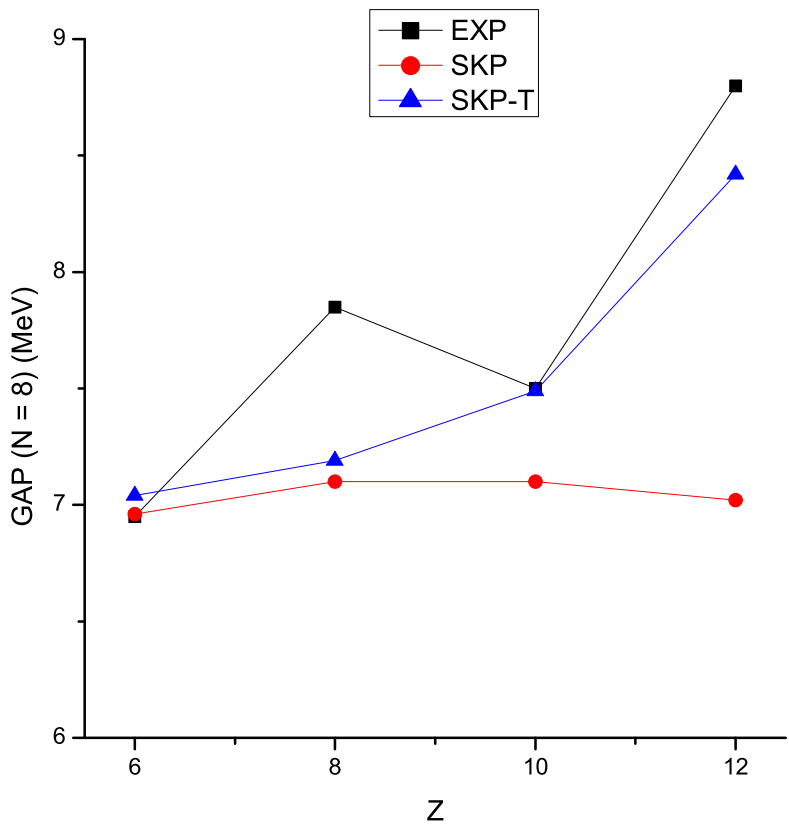

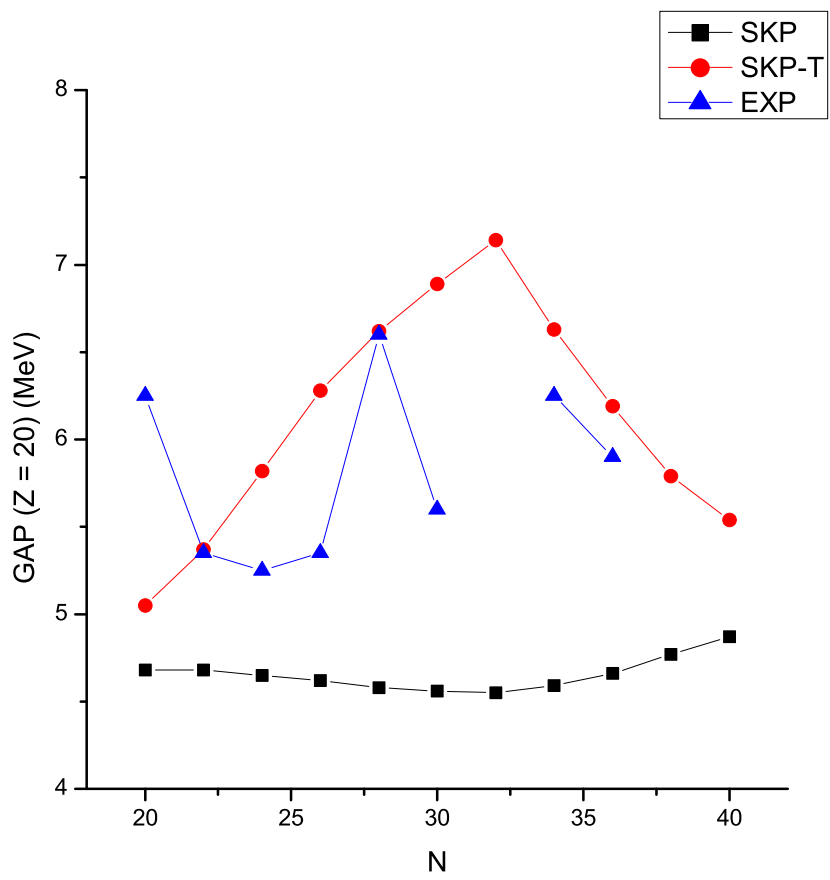

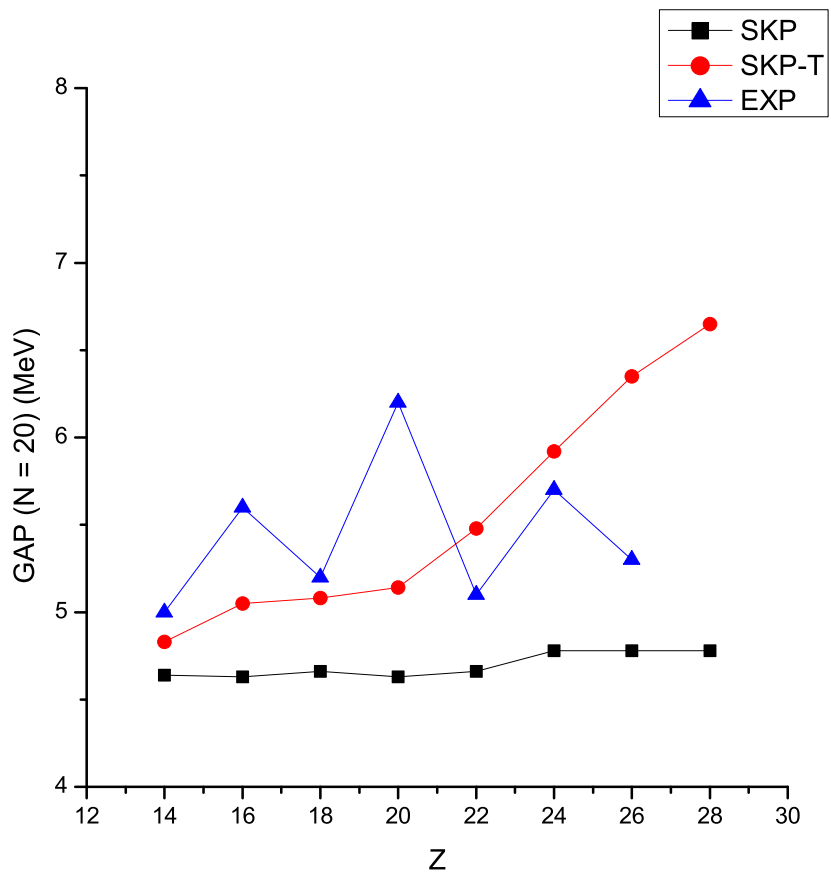

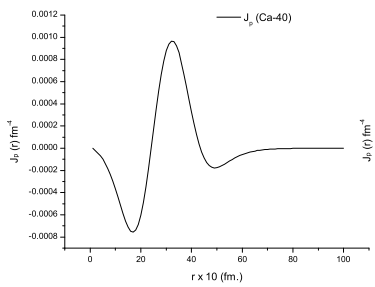
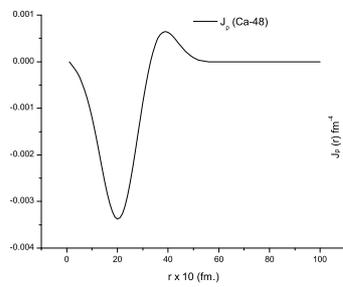
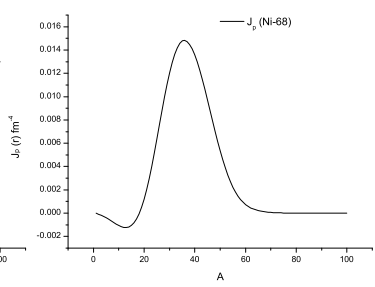
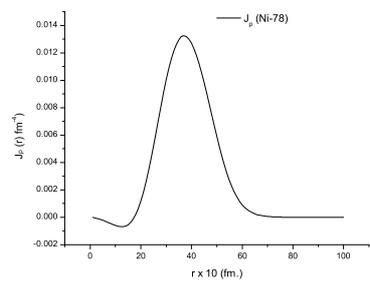
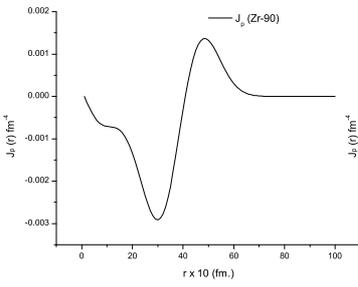
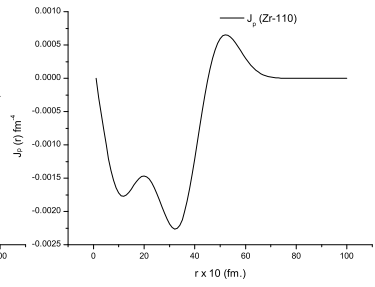
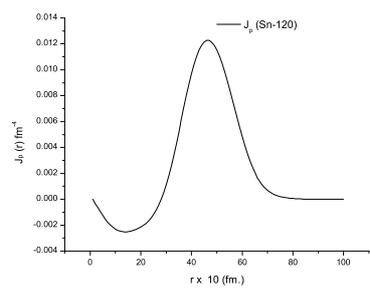
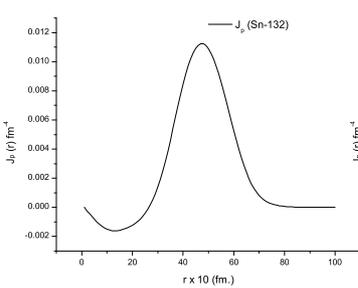
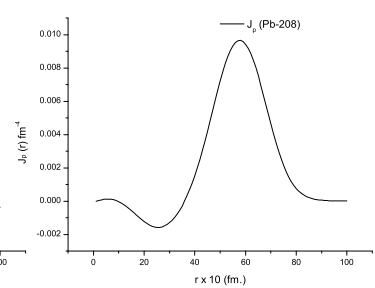

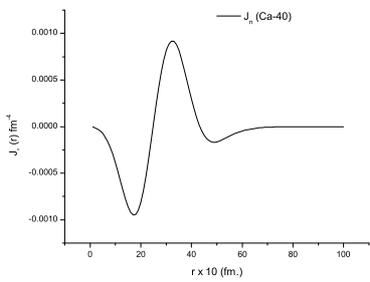
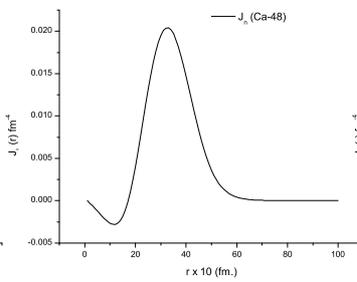
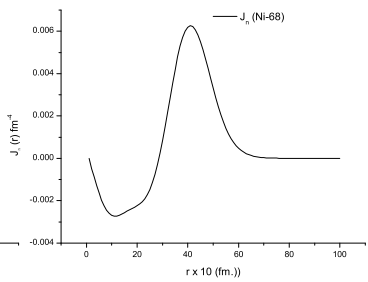
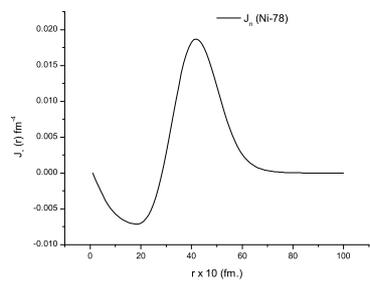
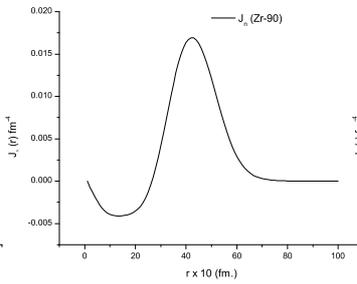
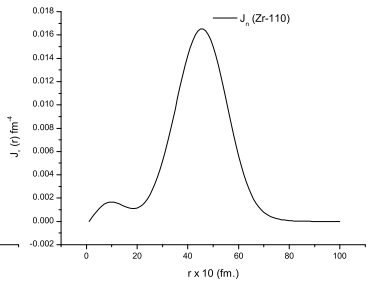
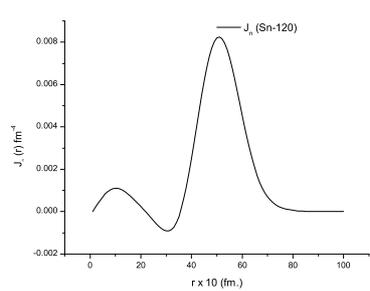
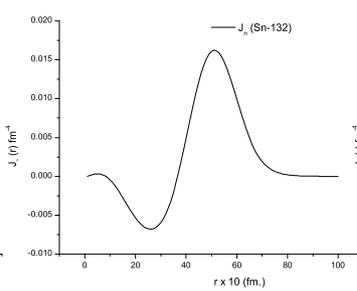
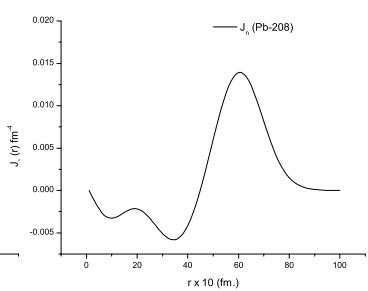